\documentclass[12pt,preprint]{aastex}

\usepackage{graphics}
\usepackage[dvips]{color}

\newcommand{\omc}{\hbox{$\omega$ Cen~}}
\newcommand{\omuno}{\hbox{$\,\omega$\,1}}
\newcommand{\omdue}{\hbox{$\,\omega$\,2}}
\newcommand{\omtre}{\hbox{$\,\omega$\,3}}
\newcommand{\omcen}{\hbox{$\omega$ Centauri~}}

\shorttitle{On the anomalous red giant branch of the globular cluster {\boldmath $\omega$}\,Cen}
\shortauthors{Freyhammer et al.}

\begin{document}
\title{ON THE ANOMALOUS RED GIANT BRANCH OF THE GLOBULAR CLUSTER {\boldmath $\omega$}\,CENTAURI
 \altaffilmark{1}
}
 \author{L.M.~Freyhammer \altaffilmark{2,3,4}, 
M.~Monelli \altaffilmark{5,6}, 
G.~Bono \altaffilmark{5}, 
P.~Cunti \altaffilmark{7}, 
I.~Ferraro \altaffilmark{5},
A.~Calamida \altaffilmark{5,6},
S.~Degl'Innocenti \altaffilmark{7,8},
P.G.~Prada Moroni \altaffilmark{7,8,9}, 
M.~Del Principe \altaffilmark{9},  
A.~Piersimoni \altaffilmark{9}, 
G.~Iannicola \altaffilmark{5}, 
P.B.~Stetson \altaffilmark{10}, 
M.I.~Andersen \altaffilmark{11},   
R.~Buonanno \altaffilmark{6}, 
C.E.~Corsi \altaffilmark{5},   
M.~Dall'Ora \altaffilmark{5,6},
J.O.~Petersen \altaffilmark{12},  
L.~Pulone \altaffilmark{5},  
C.~Sterken \altaffilmark{3,13}, and 
J.~Storm \altaffilmark{11}} 

\altaffiltext{1}
 {Based on observations made with the ESO Telescopes at the La Silla 
  and Paranal Observatories, programme IDs: 63.L-0686, 64.N-0038, 68.D-0545}
  \altaffiltext{2}
    {Royal Observatory of Belgium, Ringlaan 3, B-1180 Brussels, Belgium}
  \altaffiltext{3}
    {Vrije Universiteit Brussel, OBSS/WE, Pleinlaan 2, B-1050 Brussels, Belgium;
   lfreyham@vub.ac.be, csterken@vub.ac.be}
  \altaffiltext{4}
    {Nordic Optical Telescope, Apartado 474, E-38700 Santa Cruz de La Palma, Spain}
  \altaffiltext{5}
    {INAF-Osservatorio Astronomico di Roma, via Frascati 33, 
     Monte Porzio Catone, Rome, Italy;  bono@mporzio.astro.it, calamida@mporzio.astro.it, 
     corsi@mporzio.astro.it, dallora@mporzio.astro.it, ferraro@mporzio.astro.it, 
     giacinto@mporzio.astro.it, pulone@mporzio.astro.it}
  \altaffiltext{6}
    {Universit\`a di Roma Tor Vergata, via della Ricerca Scientifica 1, 00133 Rome, 
     Italy; buonanno@mporzio.astro.it, monelli@mporzio.astro.it}
  \altaffiltext{7}
    {Dipartimento di Fisica "E. Fermi", Univ. Pisa, Largo B. Pontecorvo 2, 56127 Pisa, Italy; 
     cunti@df.unipi.it, prada@df.unipi.it, scilla@df.unipi.it}
  \altaffiltext{8}
    {INFN, Sez. Pisa, via Largo B. Pontecorvo 2, 56127 Pisa, Italy}  
  \altaffiltext{9}
    {INAF-Osservatorio Astronomico di Teramo, via M. Maggini,
     64100 Teramo, Italy; milena@te.astro.it, piersimoni@te.astro.it}
  \altaffiltext{10}
     {Dominion Astrophysical Observatory, Herzberg Institute of
      Astrophysics, National Research Council, 5071 West Saanich Road, Victoria,
      British Columbia V9E 2E7, Canada; Peter.Stetson@nrc-cnrc.gc.ca} 
  \altaffiltext{11}
    {Astrophysikalisches Institut Potsdam, An der Sternwarte 16, D-14482 
     Potsdam, Germany; mandersen@aip.de, jstorm@aip.de}
  \altaffiltext{12}
    {Astronomisk Observatorium, Niels Bohr Institutet for Astronomi, 
     Fysik og Geofysik, Juliane Maries Vej 30, 2100 K{\o}benhavn {\O}, 
     Denmark; oz@astro.ku.dk}
  \altaffiltext{13}
    {Research Director, Belgian Fund for Scientific Research (FWO)}
\date{\centering drafted \today\ / Received / Accepted }

\begin{abstract}
We present three different optical and near-infrared (NIR) data sets 
for evolved stars in the Galactic Globular Cluster \omcen.  
The comparison between observations and homogeneous sets of stellar 
isochrones and Zero-Age Horizontal Branches provides two reasonable 
fits. Both of them suggest that the so-called anomalous branch has 
a metal-intermediate chemical composition ($-1.1 \le [{\rm Fe/H}] \le -0.8$) 
and is located $\sim 500$ pc beyond the bulk of \omc\ stars. These 
findings are mainly supported by the shape of the subgiant branch in four 
different color-magnitude diagrams (CMDs). The most plausible fit requires a 
higher reddening, $E(B-V)=0.155$ vs. 0.12, and suggests that the anomalous 
branch is coeval, within empirical and theoretical uncertainties, to 
the bulk of \omc\ stellar populations. This result is supported 
by the identification of a sample of faint horizontal branch stars 
that might be connected with the anomalous branch.
Circumstantial empirical evidence seems to suggest that the stars 
in this branch form a clump of stars located beyond the cluster.    
\end{abstract}


\maketitle

\section{Introduction}

The peculiar Galactic Globular Cluster (GGC) \omcen\ (NGC\,5139) is 
currently subject to substantial observational efforts covering 
the whole wavelength spectrum. This gigantic star cluster, the most 
massive known in our Galaxy, has (at least) three separate stellar 
populations with a large undisputed spread in age, metallicity 
(Fe, Ca) and kinematics \citep[e.g. ][]{hilker00, ferraro02, smith04}.
According to recent abundance measurements based on 400 medium resolution 
spectra collected by \cite{hilker04}, it seems that the spread in metallicity 
among the \omc\ stars is of the order of one dex  
($-2\lesssim[{\rm Fe/H}]\lesssim -1$). The metallicity distribution shows three 
well-defined peaks around ${[\rm Fe/H}]=-1.7$, $-1.5$ and $-1.2$ together with a few  
metal-rich stars at $\sim -0.8$. On the basis of high-resolution spectra, 
it has been suggested by \cite{pancino04} that stars in the anomalous branch 
\citep{lee99} might be more metal-rich, with a mean metallicity 
$[{\rm Fe/H}]\sim-0.5$. A similar metal-rich tail was also detected by 
\cite{nor96} and by Suntzeff \& Kraft (1996, and references therein). 
In the absence of conformity in the literature for the \omc\ RGB names,
we here refer to the metal-poor component as \omuno, to the 
metal-intermediate as \omdue, and to the anomalous branch as \omtre. 

Results from new observations seem to pose as many new questions as they
answer. It has been suggested \citep[e.g. ][]{lee99, hughes04} that 
the \omtre\ population might be significantly younger than \omdue. 
However, in a recent investigation based on Very Large 
Telescope (VLT) and Hubble Space Telescope (HST) data, \cite{ferraro04} 
found that the \omtre\ population is at least as old as the \omdue\ 
population and probably a few Gyrs older.    
The observational scenario was further complicated by the results 
brought forward by \cite{bedin04} on the basis of multi-band HST 
data. They not only confirmed a bifurcation along the Main Sequence, but 
also found a series of different Turn-Offs (TOs) and sub-giant branches (SGBs). 
The proposed explanations in the literature for these peculiar stellar populations 
in \omc\ are many and include, e.g., increased He content, or a separate
group of stars located at a larger distance \citep{ferraro04, bedin04, norris04}. 
The latter hypothesis could be further supported by the occurrence of a tidal 
tail in \omc\, but we still lack a firm empirical detection 
\citep[see ][]{leon00,law03}. 
The formation history and composition of \omc\ thus form a complex puzzle 
that is being slowly pieced together by investigations based on the latest generation 
of telescopes. A well-known promising technique is to study evolved stars 
simultaneously in optical and NIR bands to limit subtle 
errors due to the absolute calibration, to crowding in the innermost regions, 
and to reddening corrections.  
The main aim of this Letter is to investigate whether different assumptions 
concerning the spread in age and in chemical composition, or differences in 
distance and in reddening, account for observed optical ($BRI$) and NIR 
($JK$) CMDs. 
%For that purpose, the observations are compared with
%a homogeneous set of evolutionary models for H and He burning phases.

%-------------------------------------------------------------------------
\section{Observations and data reduction}

Near-infrared $J,K_s$ images of \omc\ were collected in 2003 with 
SOFI at the New Technology Telescope of ESO, La Silla. 
The seeing conditions were good and range from 0\farcs6 to 1\farcs1. 
Together with additional data from 2001, available in the ESO archive, 
we end up with a total NIR sample of 92 $J$- and 135 $K_s$-band images 
that cover a $14\times14$\,arcmin$^2$ area centred on the cluster. 
These data were reduced with {\tt DAOPHOTII/ALLFRAME}  following the 
same technique adopted by \cite{dallora04}.  The NIR catalogue includes 
$\sim$$1.4\times10^5$ stars. 
With FORS1 (standard-resolution mode) on the ANTU/VLT telescope, we also 
collected optical $UVI$ images in 1999. These data are from a $2\times2$ 
pointing mosaic centred on the cluster that covers an area slightly 
larger than $13\times13$\,arcmin$^2$. Exposure times range from 10 ($I$) to 
30 $s$ ($U$) and seeing conditions were better than 1\farcs0. These data 
have also been reduced using {\tt DAOPHOTII/ALLFRAME} and the photometric 
catalogue includes $\sim$$5\times10^5$ stars. Optical 
$F435W$ and $F625W$ (hereinafter $B$ and $R$ bands) data were retrieved 
from the HST archive. These images were collected with the ACS instrument 
in 9 telescope pointings, each of which provides a $BR$-image pair 
with exposure times of 12 ($B$) and 8 $s$ ($R$). The field  covered by these 
data is $\sim$$9\times9$\,arcmin$^2$, centred on the cluster. The ACS data 
were reduced with {\tt ROMAFOTwo} and provide $BR$ photometry for 
$\sim4\times10^5$ stars. The photometry was kept in the Vega system (see e.g., 
http://www.stsci.edu/hst/acs/documents). 
A detailed description of observations and data reduction will be given  
in a future paper. Here we only wish to mention that raw frames were 
prereduced using standard IRAF procedures and, in addition, the FORS1 images 
were corrected for amplifier cross-talk by following \cite{freyhammer01}.  
To improve the photometric accuracy, carefully chosen selection criteria were applied 
to pinpoint a large number ($>$100) of point-spread function stars across the 
individual frames, and several different reduction strategies were used to 
perform the photometry over the entire data set. 
The observed fields were combined to the same geometrical system using
{\tt iraf.immatch} and {\tt DAOMATCH/DAOMASTER}.  
The absolute photometric calibration of ground-based instrumental magnitudes 
was performed using standard stars observed during the same nights. The 
typical accuracy is 0.02$\,$--0.03$\,$mag for both optical and NIR data.

%--------------------------------------------------------------------------------
\section{Results and Discussion}

Figure 1 shows CMDs based on the five different photometric bands. 
The plotted stars were selected from individual catalogues by using the 
`separation index' {\em sep} introduced by \cite{stetson03}, since crowding 
errors dominate the photometric errors. The adopted {\em sep} ranges from 6.5 
(NIR data) to 8 (ACS data), which corresponds to stars having less than 
0.3\% of their measured light contaminated by neighbour stars; the higher 
the {\em sep} for a star, the less the severity of the crowding by neighbours. 
Error bars plotted at the left side of each panel account for photometric 
and calibration errors in magnitude and color. 
The input physics adopted in our evolutionary code has been discussed in 
detail in a series of papers \citep{cariulo04,cassisi98}. Here, we point 
out that the stellar models (partly available at 
http://astro.df.unipi.it/SAA/PEL/Z0.html) account for atomic diffusion, 
including the effects of gravitational settling, and thermal diffusion 
with diffusion coefficients given by \cite{thoul94}. 
The amount of original He is based on a primordial He abundance $Y_{\rm P}=0.23$ 
and a He-to-metals enrichment ratio of $\Delta Y/ \Delta Z  \sim 2.5$ 
\citep{pagel98,castellanidegl99}. 
We adopt the solar mixture provided by \cite{noels93}. For details on the 
calibration of the mixing length parameter and on model validation, we refer 
to \cite{cariulo04} and \cite{castellani03}. 
To avoid deceptive uncertainties in the comparison between theory and observations, 
the predictions were transformed into both the $BR$ Vega system and the $IJK$ 
Johnson-Cousins bands by adopting the atmosphere models provided by 
\cite{castelli97} and \cite{castelli99}.

Data plotted in Fig. 1 display the comparison between observations and a set of four 
isochrones (solid lines) with the same age (12 Gyr) and different chemical compositions 
(see labels). 
A mean reddening of $E(B-V)=0.12$ ($E(B-V)=0.11\pm0.02$, \cite{lub02})  
was adopted and a true distance modulus of $\mu=13.7$ 
($(m-M)_V=14.05\pm0.11$; \cite{thompson01}, $(m-M)_K=13.68\pm0.07$, 
Del Principe et al. 2004, private communication).
Extinction parameters for both optical and NIR bands have been estimated 
using the extinction model of \cite{cardelli89}.  
The metal-poor and the metal-intermediate isochrones supply, 
within current empirical and theoretical uncertainties, a good 
fit to the bulk of RGB and HB stars. Close  to the RGB tip, the 
isochrones are slightly brighter than the observed stars, which is
caused by the adopted mixing-length parameter ($\alpha=2$). The SGB 
and the lower RGB are only marginally affected by this parameter, 
since the empirical isochrone calibration is based on these 
evolutionary phases \citep{cariulo04}.  
However, the most metal-rich isochrone appears to be systematically 
redder than the stars at the base of the \omtre\ branch\footnote{Stars 
of the \omtre\ branch have been selected following the original detection
in the $B-R$,$R$ plane by Ferraro et al. (2004, see their Fig. 2).} (green dots) 
and fainter than the SGB stars. Clearly, the shape of the \omtre\  SGB 
does not support metal-abundances $\ge Z=0.004$, since TO stars for 
more metal-rich populations become brighter than SGB stars.
To further constrain this evidence, Fig. 2 shows the comparison between 
observed HB stars (the entire sample includes more than 2,300 objects) and 
predicted Zero Age Horizontal Branches (ZAHBs) with a progenitor age 
of 12 Gyr and different chemical compositions. 
For the sake of the comparison, the objects located between hot HB stars and 
RGB stars were selected in the $B-R$, $B$ plane and plotted as blue dots. 
Theoretical predictions plotted in this figure show that the more metal-rich 
ZAHB appears to be systematically brighter than the observed HB stars for 
$B-J\approx B-K\approx 0.5$. Moreover, the same ZAHB crosses the RGB region, 
thus suggesting that the occurrence of HB stars more metal-rich than Z=0.002 
should appear as an anomalous bump along the RGB. 
Data plotted in these figures disclose that {\em the stellar populations 
in \omc\ cannot be explained with a ranking in metal abundance}. 
In line with \cite{ferraro04} and \cite{bedin04},  we find that 
plausible changes in the cluster distance, reddening, age, and 
chemical composition do not supply a reasonable simultaneous fit of 
the \omuno, \omdue, and \omtre\ branches.

\begin{figure}
\resizebox{0.7\hsize}{!}{\includegraphics*[]{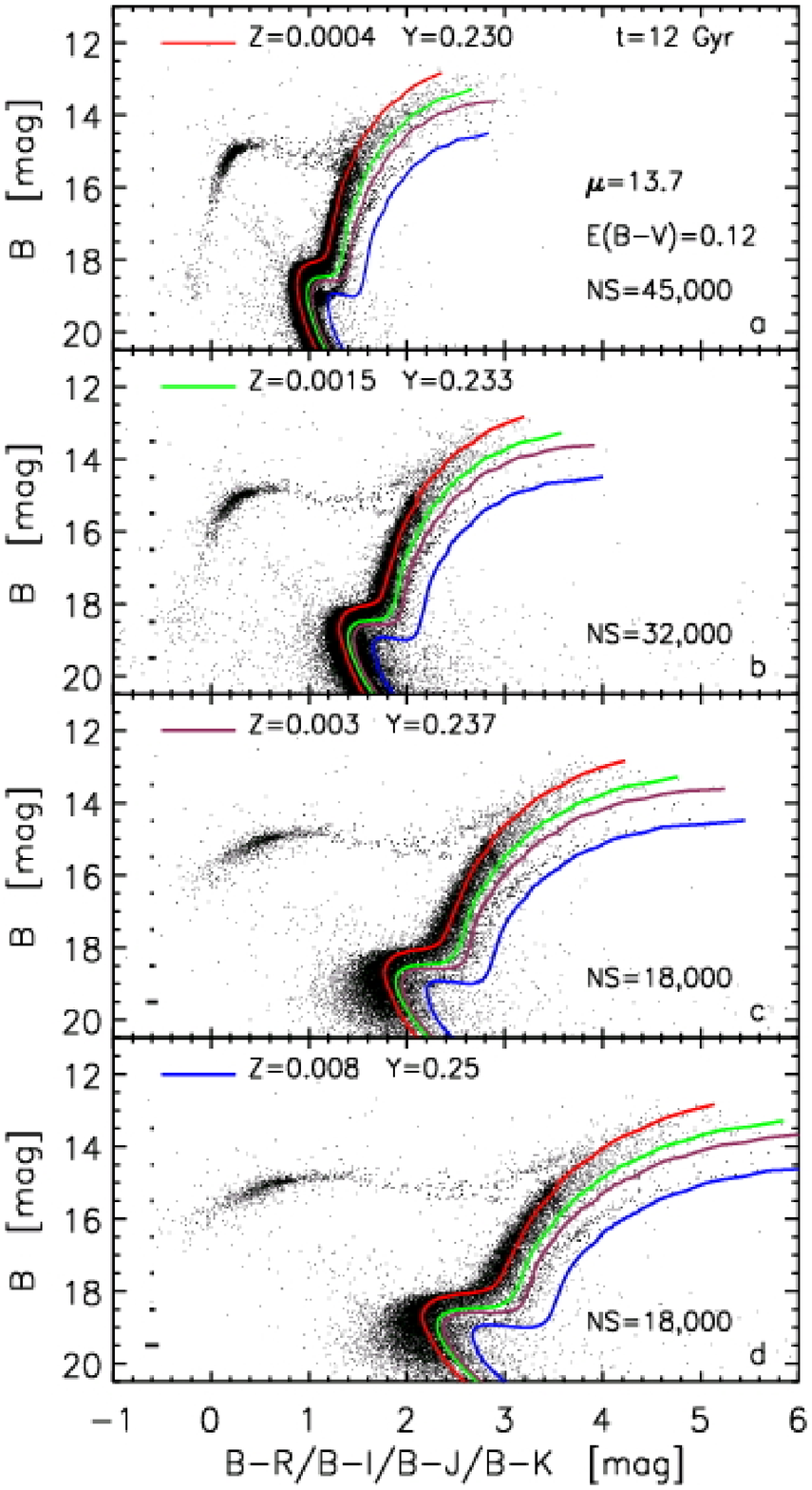}}
\caption[]{Optical (panels {\bf a,b}), and NIR (panels {\bf c,d}) CMDs for selected subsamples of the 
detected stars, compared with a set of 12 Gyr isochrones (solid lines) at different chemical 
compositions (see color coding). The adopted true distance modulus and cluster reddening are 
$\mu=(m-M)_0=13.7$ and $E(B-V)=0.12$, respectively. The number of stars selected (NS) is 
also indicated.}\label{fig:apjfig1}
\end{figure}

To further investigate the nature of the \omtre\ branch, we performed 
a series of tests by changing the metallicity, the cluster age, and 
the distance. A good fit to the anomalous branch is possible by adopting 
the same reddening as in Fig.\,\ref{fig:apjfig1}, a true distance modulus 
of $\mu=13.9$, and an isochrone of 15.5 Gyr with $Z=0.0025$ and $Y=0.248$. 
Figure\,\ref{fig:apjfig3}  shows that these assumptions supply a good 
fit in both the optical and the NIR bands, and indeed the current isochrone 
properly fits the width in color of the sub-giant branch and the shape 
of a good fraction of the RGB. Moreover, the ZAHB for the same chemical 
composition (dashed line) agrees quite well with the faint component 
of HB stars. Note that more metal-poor ZAHBs at the canonical distance 
do not supply a reasonable fit of the yellow spur stars with $15 \le B \le 15.5$. 
The yellow spur is visible in all planes, and in the NIR ({\bf c},{\bf d}) it 
even splits up in brightness due to a stronger sensitivity to effective 
temperature. The identification of the entire sample was checked on 
individual images and, once confirmed by independent measurements, 
the data indicate a separate HB sequence for the \omtre\ population. 

Although the \omtre\ fit may appear good, it implies an increase in distance 
of $\Delta\mu=0\fm2$ and an unreasonable $\sim 4$ Gyr increase in age. This estimate 
is at variance with the absolute age estimates of GGCs \citep{gratton03} and 
with CMB measurements by W-MAP \citep{bennet03}.  The discrepancy becomes even  
larger if we account for the fact that this isochrone was constructed by 
adopting a He abundance slightly higher ($Y=0.248$, vs 0.238) than estimated 
from a He-to-metals enrichment ratio $\Delta Y/\Delta Z=2.5$. This increase 
implies a decrease in age of $\sim$ 1 Gyr. Moreover, we are performing a 
differential age estimate, and therefore if we account for 
uncertainties: in the input physics of the evolutionary models (e.g. equation 
of state, opacity); in the efficiency of macroscopic mechanisms (like diffusion); 
in model atmospheres applied in the transformation of the models into the observational 
plane; and in the extinction models, we end up with an uncertainty of 
$\sim 1-2$ Gyr in cluster age \citep{castellanidegl99,krauss99}.

Owing to the wide range of chemical compositions and stellar ages adopted in 
the literature for explaining the morphology of the \omtre\ branch, we decided 
to investigate whether different combinations of assumed values of distance, 
chemical composition, and reddening may also simultaneously account for the \omtre\ 
branch and the HB stars. We found that two isochrones of 13 Gyr for $Z=0.0015$ 
and $Z=0.003$ bracket the \omtre\ stars (see Fig.\,4), within empirical 
and theoretical uncertainties. The fit was obtained using the same 
true distance modulus adopted in Fig.\,\ref{fig:apjfig3}, together with a 
mild increase in reddening, $E(B-V)=0.155$. Once again theory agrees 
reasonably well with observations in all color planes. 
Moreover, data plotted in Fig. 5 show that the predicted ZAHB with Z=0.0015 
and Z=0.003, for the adopted distance modulus and reddening correction, account 
for the yellow spur stars. The same figure shows a sample of 53  RR Lyrae stars 
selected from the variable-star catalogue by \cite{kaluzny04} for which we have 
a good coverage of $J$- and $K$-band light curves (Del Principe et al. 2004, 
private communication).
The RR Lyrae stars only account for a tiny fraction of the yellow-spur stars.
This finding together with the detection a well-defined sequence in panels
{\bf b},{\bf c}, and {\bf d} indicate that this spur might be the
HB associated with the \omtre\ population. Note that a few of these `\omtre-HB'
spur stars have also been detected by Rey et al. (2004, see their Fig. 7, for
$V\approx 14.75$) and by Sollima et al. (2004a, see their Figs. 6 and 12).

\section{Final remarks}

We have presented a new set of multi-band photometric data for the GGC \omc\ and---in 
agreement with previous findings in the literature---we find no acceptable 
fit to the different stellar populations for a single distance, reddening, and age. 
We found two reasonable fits for the \omtre\ stars: (1) by adopting a 0.2 higher  
distance modulus ($\approx$ 500 pc), a metal-intermediate composition ($Z=0.0025$, 
${\rm Fe/H}\approx -0.9$), and an unreasonable increase in age of $\sim$4 Gyr; or (2) 
for the same $\Delta\mu=0.2$ shift, an increase in the reddening, metal-intermediate 
chemical compositions ($0.0015 \le Z \le 0.003$, $-1.1 \le {\rm Fe/H} \le -0.8$), and an 
age that, within current uncertainties, is coeval with the bulk of the \omc\ stars.   
Current findings indicate that \omtre\ stars are not significantly more metal-rich 
than $Z=0.003$. This evidence is supported by the shape of the \omtre\ SGB, as 
already suggested by \cite{ferraro04} and by the fit of HB stars.  We are in favour 
of the latter solution for the following reasons:  

$\bullet$ The difference in distance between the \omtre\ branch and the bulk of 
\omc\  stars is of the order of 10\%. This estimate is  3--4 times smaller 
than the estimate by \cite{bedin04} and in very good agreement with the distance 
of the density maxima detected by \cite{odenkirchen03} along the tidal tails of 
the GGC Pal~5.  Moreover, recent N-body simulations (Capuzzo Dolcetta 
et al. 2004) indicate that clumps along the tidal tails can approximately include 
10\% of the cluster mass.     

$\bullet$ We found that by artificially shifting the \omtre-branch stars to 
account for the assumed difference in distance and reddening, they overlap with 
the \omdue\ population. It has been recently suggested by Piotto et al. (2004), 
on the basis of low-resolution spectra, that the bluer main sequence detected by 
Bedin et al. (2004) is more metal-rich than the red main sequence. Unfortunately,
current ACS photometry is only based on shallow ACS exposures, and therefore
we cannot properly identify in our data these stellar populations located in
the lower main sequence. The same outcome applies to the suspected extremely-hot
HB progeny of the bluer main sequence, since they have not been detected in the
NIR bands.  

$\bullet$ Current preliminary findings support recent N-body simulations by 
Chiba \& Mizutani (2004) and by Ideta \& Makino (2004). In particular, the 
latter authors found, by assuming that the progenitor of Omega Centauri is 
a dwarf galaxy, that more than 90\% of its stellar content was lost during 
the first few pericenter passages (see their Fig. 2).  

$\bullet$ \omc\ reddening estimates in the literature cluster around 
$E(B-V)=0.12\pm0.02$. However, the map from \citep{schlegel98} indicates 
reddening variations of 0.02 across the body of the cluster while, more 
importantly, 2MASS data \citep{law03} show a very clumpy reddening 
distribution at distances beyond 1$^\circ$ (100$\,$pc) from the cluster 
centre, with large variations $\Delta E(B-V)=0.18$ across a 4\,degrees$^2$ 
area. However, the inference of reddening {\it between\/} the main 
body of \omc\ and the supposed background population is very surprising.  
\omc\ lies at the comparatively low Galactic latitude of $+15^\circ$, 
and the reddening variations seen in the Schlegel and 2MASS maps likely 
originate in the foreground interstellar material of the Galactic disk.  
Any interstellar material behind \omc\ must lie at least 1.4$\,$kpc 
from the Galactic plane, and therefore would most likely be associated 
with \omc\ itself.  Smith et al. (1990) have reported a significant
detection of $H\,I$ in the direction of \omc, blueshifted by
$\sim\,40\,$km s${}^{-1}$ with respect to the cluster velocity. 
However, their interpretation is that this gas is associated with the 
northern extension of the Magellanic Stream far beyond the cluster. 
It would be a remarkable coincidence if this interstellar material 
happened to lie between the cluster and a clump in the tidal tail 
at a 10\% greater distance than the main cluster body, while traveling 
at the quoted relative speed. However, it is worth noting that 
Smith et al. (1990) estimated for this cloud a column density of 
$N_H\approx 3\times10^{18}$  atmos/cm$^2$, and therefore a reddening 
$E(B-V)\approx0.07$ (Predehl \& Schmitt 1995) that is at least 
a factor of two larger than required by our best fits. This 
point is crucial for the proposed explanation and needs to be 
further investigated.     
 
Finally, we note that the comparison between predicted ZAHBs and 
HB stars indicates that the occurrence of an old stellar population 
with $Z > 0.002$ would imply the occurrence of an anomalous clump 
along the RGB. In fact, more metal-rich,
red HB stars cover the same color range of metal-poor RGB stars. The
detection of such a feature along the RGB can supply robust constraints
on the progeny of the \omtre\ stellar population. 
These results, when independently confirmed, would suggest 
that the \omtre\ branch might be a clump of stars located 500 pc beyond 
the bulk of the cluster. No firm conclusion can be drawn on the basis 
of current data, although this evidence together with the increase in 
radial velocity among \omtre\ stars measured by \cite{sollima04} and numerical 
simulations recently provided by Capuzzo Dolcetta et al. (2004) indicates that 
it could be a tidal tail.  

\section{Acknowledgments}
It is a pleasure to thank M. Sirianni and N. Panagia for fundamental 
suggestions on the calibration of ACS data. We wish to tank two 
anonymous referees for their suggestions and pertinent comments that 
helped us to improve the content and the readability of the manuscript.  
We are also grateful to V. Castellani for his critical reading of an 
early version of this manuscript. 
This work was supported by the Belgian Fund for Scientific Research (FWO) 
in the framework the project ``IAP P5/36`` of the Belgian Federal Science Policy,   
by MIUR/PRIN~2003 in the framework of the project: ``Continuity and Discontinuity 
in the Galaxy Formation``, and by Danish NSRC in the framework of the project: 
``Structure and evolution of stars: new insight from eclipsing binaries and 
pulsating stars``. 

%--------------------------------------------------------------------------------------

\clearpage

%-------------------------------
%

%-----------------------------------------------
%
\begin{figure}
\resizebox{0.7\hsize}{!}{\includegraphics*[]{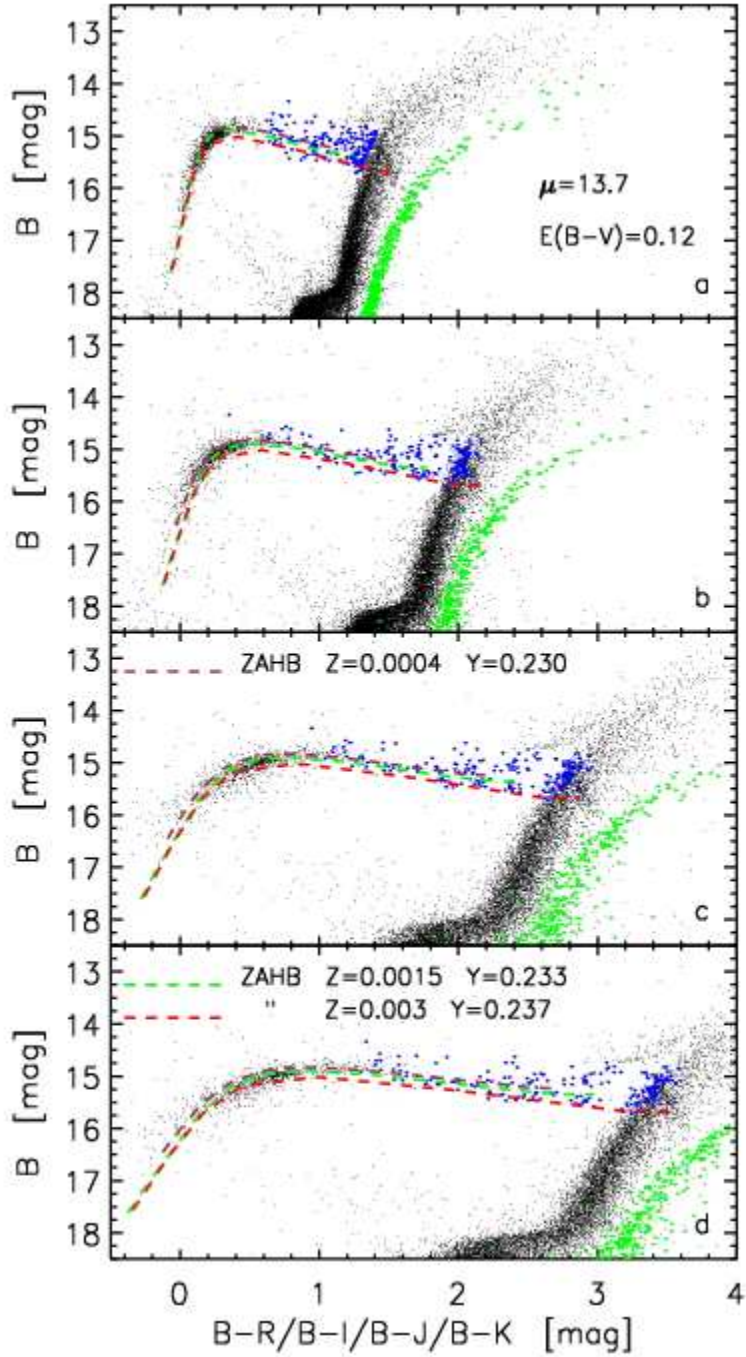}}
\caption[]{Same as in Fig.\,\ref{fig:apjfig1}, but the comparison between theory 
and observations is focused on Horizontal Branch stars. Blue objects mark stars 
located between hot HB stars and RGB stars. They have been selected in $B-R$,$B$ 
plane.}\label{fig:apjfig2}
\end{figure}
%-----------------------------------------------
%
\begin{figure}
\resizebox{0.7\hsize}{!}{\includegraphics*[]{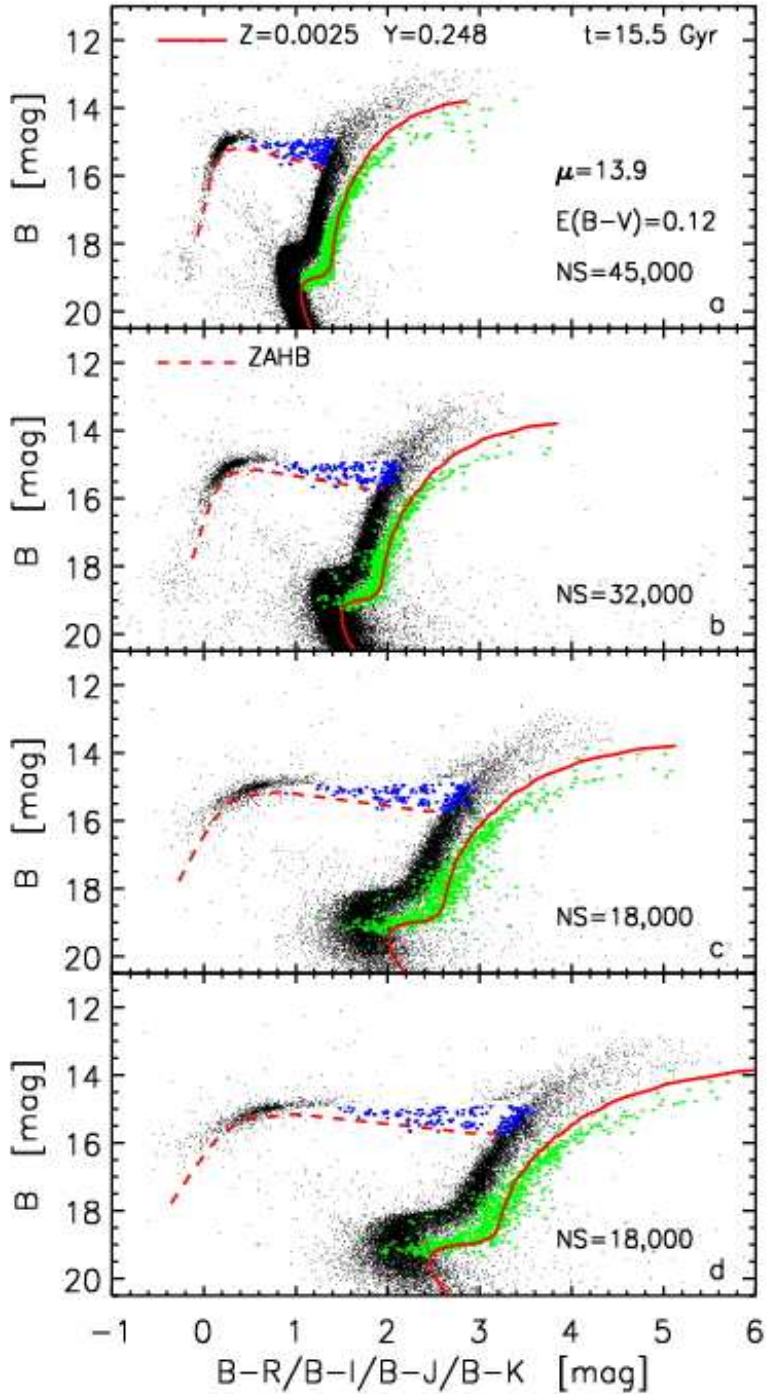}}
\caption[]{Same as in Fig.\,\ref{fig:apjfig1}, but compared to a 15.5 Gyr isochrone 
for the \omtre\ population, constructed by adopting $Z=0.0025$, and  $Y=0.248$. The 
fit was performed by adopting a true distance $\mu=13.9$, and a reddening correction 
$E(B-V)=0.12$. Note that the corresponding ZAHB matches the selected HB stars 
(blue objects).} \label{fig:apjfig3}
\end{figure}
\begin{figure}
\resizebox{0.7\hsize}{!}{\includegraphics*[]{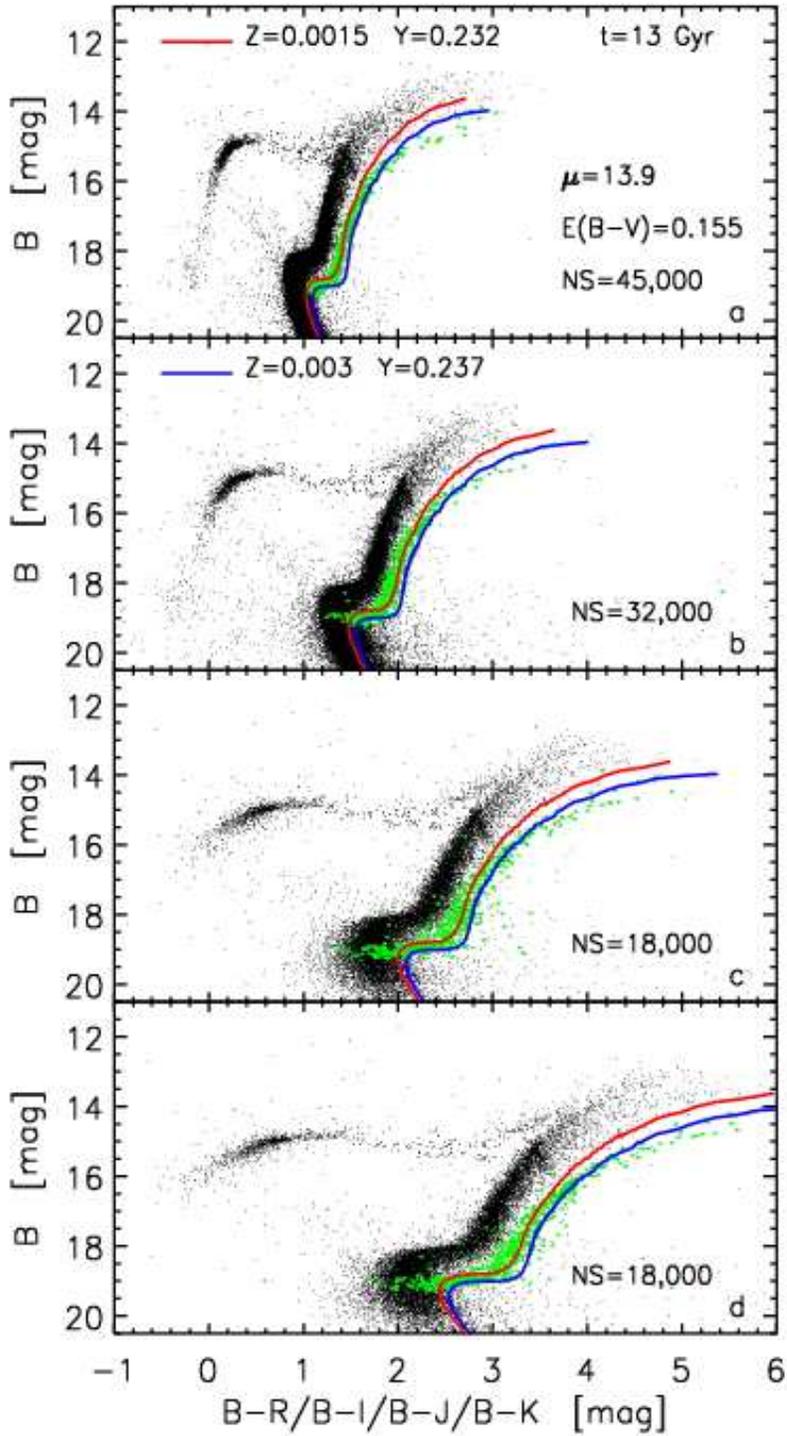}}
\caption[]{Same as Fig.\,\ref{fig:apjfig3}, but compared to 13-Gyr isochrones 
constructed by adopting different chemical compositions (see labels) and a higher 
reddening.}\label{fig:apjfig4}
\end{figure}
%-----------------------------------------------------------------------------------------

\begin{figure}
\resizebox{0.65\hsize}{!}{\includegraphics*[]{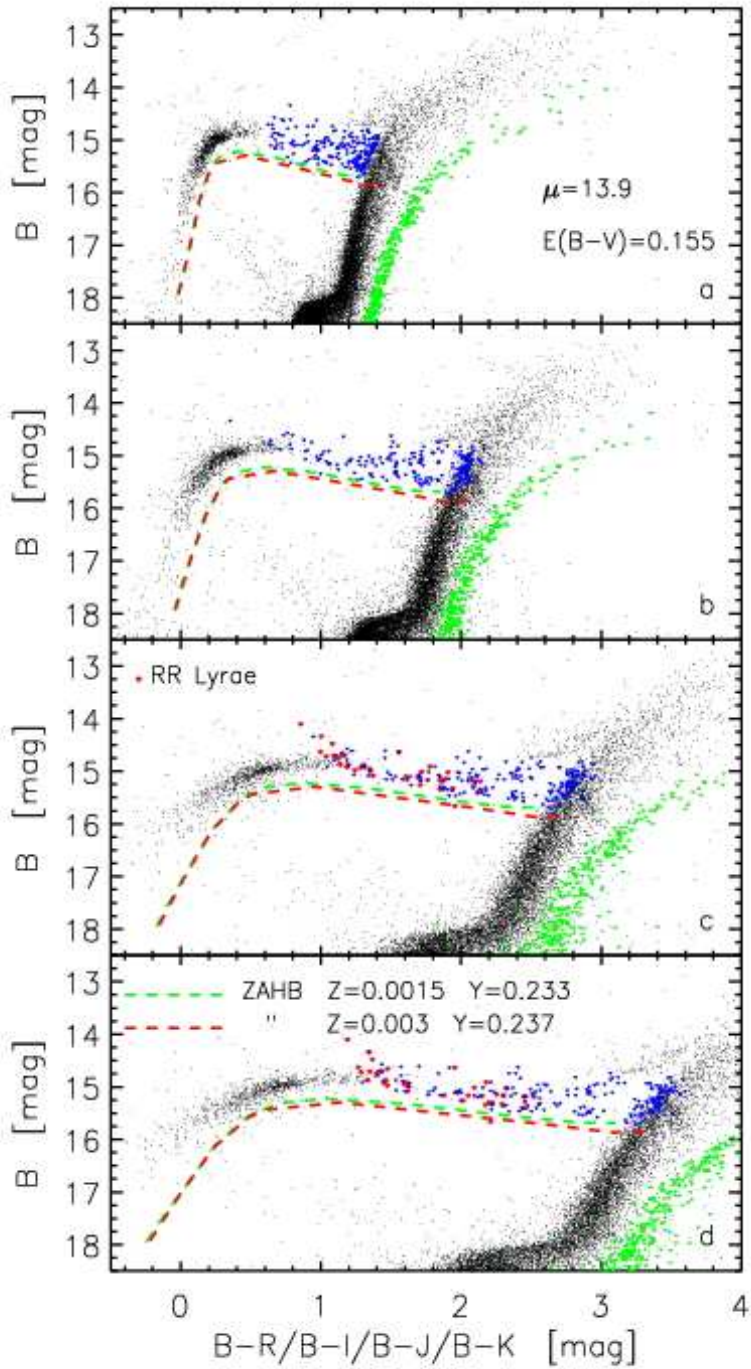}}
\caption[]{Same as in Fig.\,\ref{fig:apjfig4}, but the comparison between theory 
and observations is focused on Horizontal Branch stars. Blue objects mark stars 
located between hot HB stars and RGB stars, while red dots (panels {\bf c,d}) in 
the HB region display RR\,Lyrae stars for which we have accurate mean NIR 
magnitudes.}\label{fig:apjfig5}
\end{figure}
%-----------------------------------------------------------------------------------------

\end{document}